# Analyzing Curricular Pattern Complexity Using AI to Improve On-Time Graduation Rates

Lynn Vonderhaar
*Department of Electrical Engineering and Computer Science*
*Embry-Riddle Aeronautical University*
Daytona Beach, USA
vonderhl@my.erau.edu

Juan Couder
*Department of Electrical Engineering and Computer Science*
*Embry-Riddle Aeronautical University*
Daytona Beach, USA
ortizcoj@my.erau.edu

Siri Siqveland
*Department of Electrical Engineering and Computer Science*
*Embry-Riddle Aeronautical University*
Daytona Beach, USA
siqvelas@my.erau.edu

Omar Ochoa
*Department of Electrical Engineering and Computer Science*
*Embry-Riddle Aeronautical University*
Daytona Beach, USA
ochoao@erau.edu

James Pembridge
*Department of Engineering Fundamentals*
*Embry-Riddle Aeronautical University*
Daytona Beach, USA
pembridj@erau.edu

*Abstract*—**The rise of Artificial Intelligence (AI) enables automatic analysis of large amounts of data. Previously time-consuming and labor-intensive tasks can be completed much more efficiently with the use of AI. This work uses AI techniques to analyze and revise curricular patterns in an undergraduate degree for Software Engineering. Curricula often have long sequences where failure to pass a class within the sequence may jeopardize completion of the degree within four years. Manual analysis and revision of curricula by university faculty is a lengthy and labor-intensive process, causing changes to occur rarely and making it impossible to keep up with the changing needs of students. This work reduces the time-to-change for curricula and reduces bottlenecks and graduation delays by using Large Language Models (LLMs) to analyze curricular patterns and suggest revisions.**

## I. INTRODUCTION

This work uses Artificial Intelligence (AI) to analyze and revise curricular patterns in a university undergraduate degree for Software Engineering. Many university curricula have long curricular patterns and sometimes have a critical path that spans most, or all of the semesters in a student's program. Long curricular patterns can jeopardize the four-year duration of undergraduate degrees. If there is a set of classes that must be taken in sequence, the chance of pushing student graduation by a semester, or even an academic year, increases significantly. Manual analysis and revision of curricula by university faculty to address these issues is a lengthy and labor-intensive process, rendering these types of changes infrequent and making it difficult to keep up with the changing needs of students.

With the advent of AI, the analysis of large amounts of data is possible, enabling automatic examination of curricular patterns and prerequisites. This paper proposes a method to analyze and revise curricular patterns, e.g., sequences of classes, to improve on-time graduation rates and graduate more prepared students. This paper leverages Large Language Models (LLMs) to analyze and revise the curricular patterns of an undergraduate degree in Software Engineering to reduce bottlenecks and graduation delays caused by outdated curricular sequences. The LLM gauges whether class prerequisites are necessary and suggests curricular changes to reduce bottlenecks within the degree program. Specifically, this paper contributes to the knowledge domain by comparing the performance of five LLMs and an LLM with Retrieval-Augmented Generation (RAG) and identifies the best tool for this analysis.

Curricular complexity and patterns have been analyzed previously where classes are modeled as nodes on a directed graph and analyzed based on structural and instructional complexity [1]. LLMs are not only faster but can also improve this method by incorporating context into the analysis and offering solutions to reduce complexity. This work will modify the curriculum for a university undergraduate degree in Software Engineering and validate the changes by surveying university faculty. The contributions of this work are as follows:

1. Comparing the capabilities of multiple LLMs and an LLM with RAG in curricular analysis.
2. Using an LLM to analyze curricular complexity for an undergraduate degree.
3. Revising curricular patterns with an LLM.
4. Surveying university faculty to validate the curricular changes.

## II. BACKGROUND

Before discussing the approach of this paper, it is critical to understand some background concepts including curricular complexity calculations, LLMs, and RAG.

### A. Curricular Complexity

Curricular complexity refers to the degree to which the structure of a degree program affects the student's progress towards graduation [2]. If many classes depend on many others in long chains, students have less flexibility when deciding what classes to take, making delays in graduation more likely. For instance, if there is a chain of four classes that need to be taken in sequence, failing class A would delay classes B-D. When several courses depend on one specific course, that course is known as a bottleneck course [3]. Curricular complexity aims to quantify the structural difficulty of a degree. If a curriculum has low complexity, there tends to be more parallel progress instead of sequential courses.

Complexity of a curriculum is calculated using the following definitions:

1. Blocking factor: the number of courses that a student is prevented from taking until a specific course is passed.
2. Delay factor: the longest chain of prerequisites that passes through a course.
3. Cruciality: the sum of the blocking and delay factors for a course.
4. Curricular complexity: the sum of the cruciality of all courses in the curriculum [1].

### B. Large Language Models

LLMs are predictive machine learning models designed to understand and generate human language. LLMs are based on the transformer architecture which allows them to understand the context of sequences of words and predict the next word in a specific sequence [4]. They are trained on extremely large datasets of texts, the more data they are trained on, the better they tend to capture the human language. Transformer-based architectures are notable for converting words into tokens, which then are converted into numerical vectors. Once the words are converted into those numerical vectors, the models can learn what words in the sentences are most important. This sequence is repeated through each of the layers of the model until the model is capable of not only knowing what words are important, but also what words follow others typically in the texts.

There is a great variety of available LLMs such as OpenAI's ChatGPT, Anthropic's Claude, Google's Gemini, Amazon's Nova, and many others [5]. Despite their power, LLMs are susceptible to several limitations such as knowledge cutoff, when they only know the information they are trained with, and more recent information is unknown to them, or hallucinations, where they can generate confident responses, even when they are incorrect [6].

### C. Retrieval-Augmented Generation

To prevent the limitations LLMs run into, RAG was introduced. RAG is a technique that combines LLMs with external knowledge retrieval systems to update the knowledge they have by feeding them more up-to-date information during its runtime [7]. The RAG process references the external knowledge base for every LLM prompt so that the LLM can gain that extra knowledge required to respond more accurately to the user. RAG reduces hallucinations by grounding its answers in real documents instead of guessing them. This also allows LLMs to have access to private or internal data without being trained on it. Another benefit is increased explainability, as the LLM can point to the exact document used to generate the answer. Fig. 1 shows the RAG process [8].

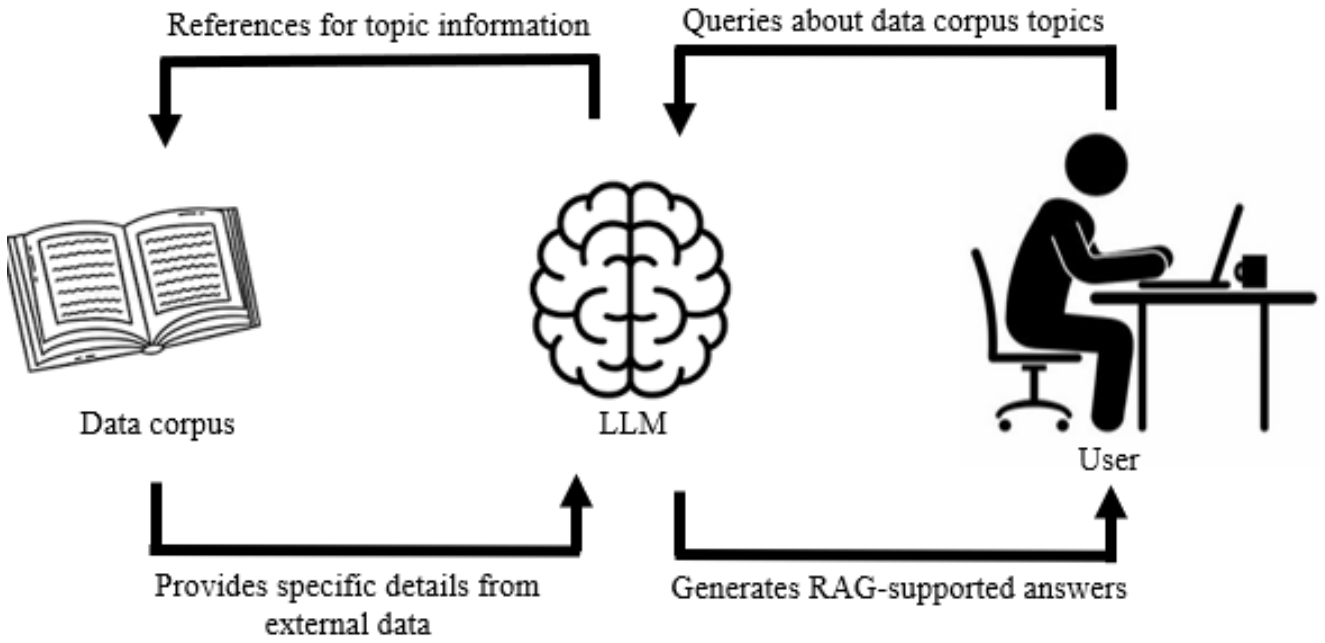

Fig. 1. The RAG process [8].

## III. Approach and Results

This work compares multiple popular LLMs, as well as an LLM with RAG to determine the best tools for analyzing and revising curricular complexity.

Each of the chats followed the same general format:

1. Provide the LLM with Heileman, et al.'s curricular complexity paper [1].
2. Ask the LLM to perform a similar analysis based on a given degree program flow chart and course descriptions.
3. Request actionable revisions based on the analysis.

The degree flow charts and course descriptions are at https://github.com/lynndalou/CurricularComplexity. Each of the LLMs or systems was tested on the B.S. in Software Engineering degree program and their performances compared. The performance comparison is shown in Table I, which summarizes each model's comprehension of the complexity concepts, the calculation of the blocking and delay factors, the data formatting, and the percentage of reasonable suggestions where "reasonable" includes suggestions that are possible, e.g., removing a non-existent prerequisite is not reasonable.

Table I. Summary of the LLM and system performances on the B.S. in Software Engineering.

| LLM/System | Description of Complexity Concepts | Complexity Calculations | Percentage of Reasonable Suggestions | Data Formatting |
|---|---|---|---|---|
| GPT 5.3 | Correct | Incorrect | 80% | Text response |
| Sonnet 4.6 | Correct | Correct | 83% | JavaScript tools |
| Gemini 3.1 Flash Lite Preview | Correct | Incorrect | 100% | Text response |
| Nova 2 Lite | Correct | Incorrect | 17% | Text response |
| Claude Opus 4.1 | Correct | Incorrect | 71% | Text response |
| Sonnet 4.6 + RAG | Correct | N/A | N/A | Text response |

### *A. Chat GPT 5.3*

The first LLM tested was Chat GPT 5.3 [9]. The full chat history from the OpenAI website can be found at https://github.com/lynndalou/CurricularComplexity.

This chat followed the chat outline described earlier in this section. GPT 5.3 correctly explained the curricular complexity concepts from the Heileman, et al. work [1]. GPT 5.3 also identified many of the main skeletons of curricular dependency. The LLM outlined the curricular bottlenecks, provided an organized table showing the cruciality of each course, and offered several suggested revisions to reduce choke points. The suggested revisions included:

1. Removing Software Engineering Practices (SE 300) as a prerequisite for Software Construction (SE 320), given that SE 320 requires programming knowledge but encapsulates all of the software engineering knowledge that it requires.
2. Removing the synchronization delay from both Analysis and Design of Software Systems (SE 310 and SE 320) being prerequisites for senior capstone (SE 450 and SE 451).
3. Moving discrete structures (CS 222) earlier in the four-year degree program.
4. Removing Computer Science II (CS 225) as a prerequisite for Operating Systems (CS 420), potentially replacing the prerequisite with Data Structures (CS 315), which has lower blocking and delay factors.
5. Introducing a new course covering software engineering foundations to be taught between basic programming and SE 300 to reduce the conceptual jump between courses and improve the SE 300 pass rate.

Although much of the analysis and suggested revisions seemed reasonable, the underlying complexity calculations were incorrect, making it impossible to fully trust the rest of the outputs.

Upon request, GPT 5.3 did offer an example model of how the suggested revisions would affect on-time graduation rates, but the LLM did note that it could not definitively model these percentages without knowing the actual pass rates of the individual courses.

### *B. Claude Sonnet 4.6*

The second LLM tested was Sonnet 4.6 [10]. The full chat history is also available on the GitHub repository. This chat followed the same outline described earlier in this section. Sonnet 4.6 correctly explained the curricular complexity concepts from Heileman, et al. [1]. Sonnet 4.6 correctly calculated the blocking factors, delay factors, and cruciality for the courses in the degree program. However, it did note the limitation that it could not calculate instructional complexity without having the pass rate data from the university [1].

Sonnet 4.6 also formatted the output in an easily understandable way by providing an interactive JavaScript tool to make the information easier to visualize and digest. A selection from this tool is shown in Fig. 2. The tool provides analysis for each of the semesters within the four-year degree program.

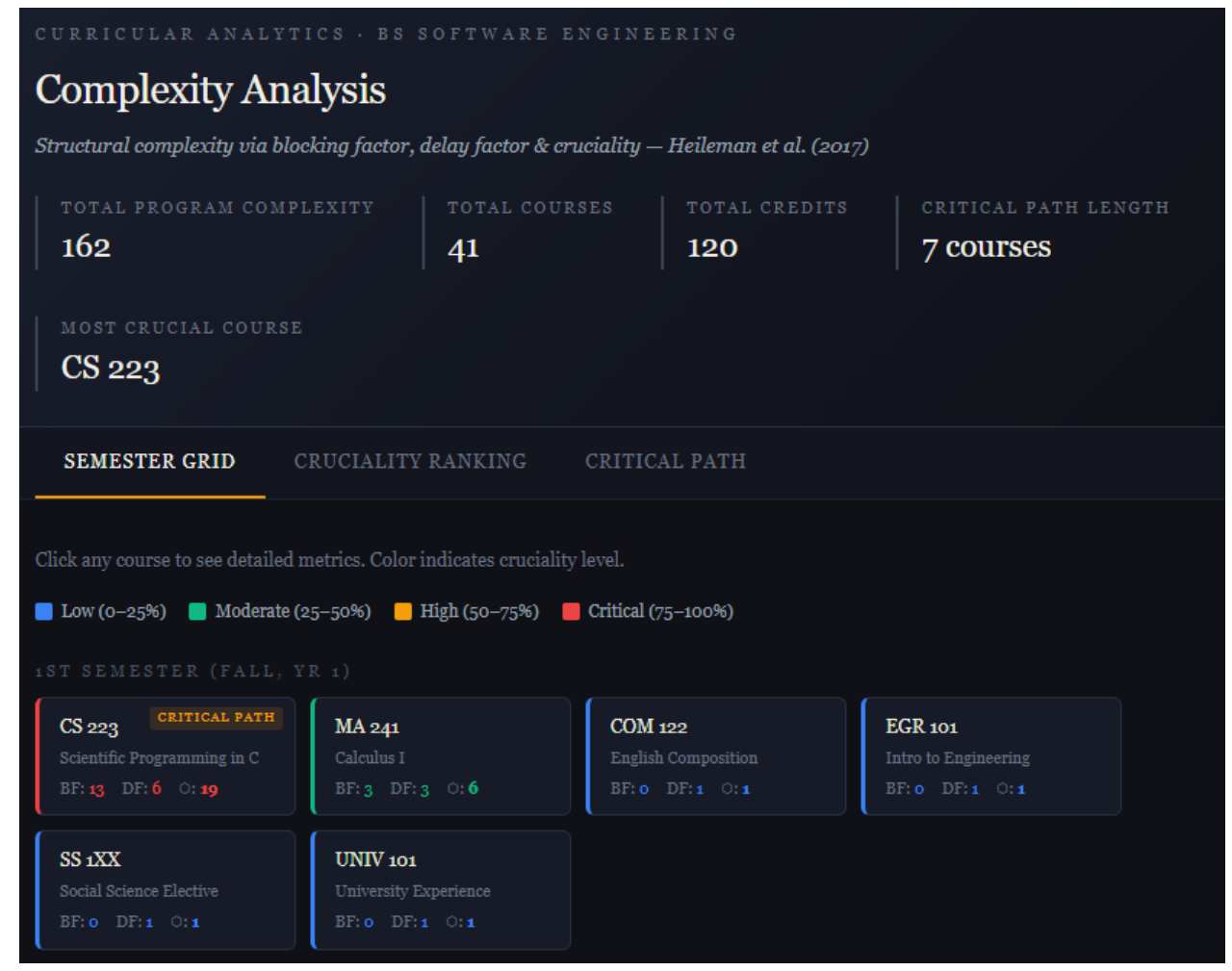

Fig. 2. A selection from the Sonnet 4.6 curriculum analysis tool.

Sonnet 4.6 provided six potential revisions and provided a formatted report explaining the findings and the suggested revisions. The LLM suggested the following six revisions:

1. Offering summer sessions for early, crucial classes, e.g., Scientific Programming in C (CS 223) and CS 225.
2. Moving CS 222 earlier in the degree program.
3. Offering SE 310 and SE 320 in both the fall and the spring semesters.
4. Offering more academic support, e.g., tutoring and supplemental instruction, for CS 223 and CS 225.
5. Either offering Computer Architecture (CEC 470) in both the fall and spring semesters or removing it altogether and incorporating crucial information for software engineers into other courses.
6. Front-loading non-math courses for students to come in at a lower-level math than Calculus I (MA 241) to reduce graduation delay resulting from that.

It is interesting to note at this point that both GPT 5.3 and Sonnet 4.6 suggested moving CS 222 to an earlier semester to start that curricular path sooner. However, Sonnet 4.6 specifically suggested moving CS 222 to the second semester, where it already is located. Additionally, many of the other suggestions revolve around offering courses in both the fall and spring semesters, which is not possible due to the size of the department. After adding this constraint, Sonnet 4.6 offered these refined suggestions:

- Allowing SE 320 to be a corequisite with senior capstone instead of a prerequisite as failure to pass SE 320 in the 6th semester would cause a full year's delay in graduation.

- Removing CEC 470 as a requirement for software engineering students.
- Moving Operating Systems (SE 420) to semester 6.
- Moving Information and Computer Security (CS 432) to semester 6.

Despite the most recent prompt explaining low enrollment in CS 225 and SE 300, the LLM repeated its suggestion that the biggest point of concern is that a student could fail CS 225 or SE 300 and be put behind a full year because they are both offered only once per year. Sonnet 4.6 urged a reconsideration of offering both courses in both the fall and spring semesters if enrollment allowed for it.

Upon request, Sonnet 4.6 provided a JavaScript tool for running the simulation of how the recommended changes would affect on-time graduation rates. This tool provided customization of the base pass rate, the percentage of underprepared students, the percentage of per semester drop-out risk, and the total number of students. This customization accounts for the LLM not having access to the precise pass rates for each of the courses.

### C. Gemini 3.1 Flash Lite Preview

Gemini 3.1 Flash Lite correctly described the concepts of curricular complexity as defined by Heileman, et al. [1, 11]. It also correctly noted that it could not calculate the instructional complexity when only given the degree flow chart and course descriptions. The LLM correctly identified the prerequisite chains and the bottlenecks but was unable to correctly calculate the blocking and delay factors.

This LLM suggested two revisions to the curriculum, which were:

- Exploring whether SE 310 and SE 320 can be taken concurrently or in earlier semesters to reduce the delay factor in senior year.
- Introducing an early bridge course combining computer science and math topics.

Gemini 3.1 Flash Lite did model the impact on graduation rates and noted its assumption of a standard 80% pass rate but did not format the information as well as Sonnet 4.6. The full chat can be found on the project GitHub.

### D. Nova 2 Lite

Nova 2 Lite did correctly describe the complexity concepts and noted that without explicit pass rates, it would need to assume default values for some calculations [12]. However, that was the extent of its performance. It did not provide correct complexity calculations. The LLM identified some correct bottleneck courses, e.g., CS 225, but added some bottlenecks where it should not have, e.g., Calculus II (MA 242), which only has a blocking factor of 1.

Nova 2 Lite also provided many poor suggestions for curricular improvement including:

- Parallelizing Calculus I and II,
- Offering Microprocessor Systems lecture and lab in the same term, which they already are, and
- Moving senior capstone courses to earlier in the degree program.

The full chat history can be found on the project GitHub.

### E. Claude Opus 4.1

Opus 4.1 correctly described the complexity concepts and noted that it would need historical pass rates in order to provide success simulations [13]. An interesting characteristic of Opus 4.1 is that it maintains its chain of reasoning for review within the chat history.

This LLM correctly identified the main prerequisite paths within the curriculum, however it consistently made mistakes in calculating the delay factor for courses, seemingly not using the full prerequisite chain in the calculation. Its suggested revisions also varied in effectiveness. For example, the LLM suggested removing CS 255 as a prerequisite for CS 222, but this prerequisite does not exist. However, it also suggested making 300 and 400 level courses less sequential, which could be a reasonable suggestion. The full chat history can be found on the project GitHub.

### F. Sonnet 4.6 + RAG

Based on the performance of each tested LLM, Sonnet 4.6 offered the most accurate and helpful responses. Therefore, Sonnet 4.6 was chosen for a follow-up experiment to see if the use of RAG would improve its performance.

In this experiment, Heileman, et al.'s paper, the B.S. in Software Engineering degree flow chart, and the course descriptions were added to the LLM's RAG data corpus [1]. The chat followed a similar outline each of the previous chats, starting from a new chat, but referenced the data corpus instead of uploading the referenced files.

Although Sonnet 4.6 with RAG was able to correctly define the blocking and delay factors, that was the extent of its capabilities when using a simple architecture with standard embedding. Even with a large chunk size and chunk overlap, the LLM often became confused by the information that it retrieved, mainly due to difficulties in retrieving the flow chart. Because the flow chart is a visual document, its full utilization would require an additional model, e.g., an Optical Character Recognition (OCR) model or a computer vision model, to improve its performance. However, the addition of a second model within the pipeline greatly reduces the accessibility of this method to a wide range of university departments, therefore making it a less viable option than using a pretrained LLM without RAG.

## IV. Discussion

The purpose of this study was to identify whether LLMs could be used to analyze curricular complexity and if so, what the best LLM for this purpose is. Through testing several LLMs, it is clear that Anthropic's Claude Sonnet 4.6 is the most effective LLM for analyzing curricular complexity due to its ability to correctly calculate the various complexity factors and because of its helpful formatting of the results.

After determining that Sonnet was the best tool for this analysis, it was further prompted for suggestions on how to change the curriculum to reduce its complexity and increase on-time graduation rates. It is important to note that this analysis took some back and forth prompting to tune the responses as it did make some unreasonable suggestions, e.g., moving CS 222 to the semester it was already in, and also made suggestions to offer more sections of classes without initially taking into account faculty constraints. The following is a list of the suggested changes:

1. **Removing SE 310, SE 320, or both as a prerequisite to senior capstone**. Since SE 310 and SE 320 are both offered only once per year, failure of even just one of them delays a student's graduation an entire year.
2. **Reviewing whether Computer Architecture (CEC 470) is necessary for software engineering students**.
3. **Moving Software Quality Assurance (SE 420) to semester 6 instead of having it in semester 8**. Sonnet 4.6 argued that its only prerequisite is in semester 3, so it does not need to be so late in the curriculum and could pose as a final semester trap.
4. **Moving Information and Computer Security (CS 432) to semester 6** to free up space in a student's final year for any courses that either were not taken earlier or for retaking any courses.
5. Sonnet's most adamant suggestion was **offering CS 225 and SE 300 in both spring and fall semesters** because they are the beginning of the curriculum's critical path and could easily set students back an entire year.

These suggested curricular revisions were validated by the two most senior software engineering faculty within the department. Their feedback is shown in Table II. The faculty feedback indicates that Sonnet's suggestions are a good starting point, but that some things may have unintended consequences. For example, Faculty Member One noted that moving Software Quality Assurance (SE 420) to the third year would affect the cross-listing of that course with the graduate-level course, which is done due to faculty and enrollment constraints. Additionally, Faculty Member One expressed concern that moving many of these courses earlier in the curriculum could cause credit overload in earlier semesters, so these suggestions could not be adopted in isolation. They would need to be adopted along with additional modifications to ensure equal credit distribution. It is also important that both faculty members noted that CS 225 and SE 300 are already offered in both the fall and spring semesters, indicating a disparity in the course catalog and flow chart information, or simply a limitation of Sonnet's understanding of the curriculum.

Though the faculty members expressed interest in each of these suggestions, they were hesitant to take them without further analysis of their effects. Therefore, before bringing them to the attention of a curriculum committee, it would be important to model their effects on graduation rates. Faculty Member Two also noted that they did not want to use the suggestions in isolation, but rather in addition to larger structural and competency changes.

## V. Related Work

The related work in this area is two-fold: literature regarding analysis of curricular complexity, and literature regarding the use of LLMs for curricular analysis.

### A. *Analyzing Curricular Complexity*

There is existing literature that utilizes modeling software to analyze the complexity of curricular patterns in engineering programs and improve on-time graduation [1, 14, 15]. Heileman, et al. model classes as nodes in a directed graph with the class order forming edges. They define curricular complexity as the delay factor, i.e., the longest chain of prerequisites to a course, and blocking factor, i.e., the courses that students are blocked from taking without passing that course, in the graph. The authors then simulate the effect of curricular changes on graduation delays. Reeping and Grote also analyze curricular complexity but do so in relation to minimizing the delay and blocking factors for transfer students [14]. Slim, et al. propose a metric called course cruciality, which measures how important a course is based on its delay and blocking factors and the number of degree programs that require the course [15]. However, none of these works incorporate AI into their analysis process, which, as this work shows, can speed up the analysis drastically.

In another paper, Wigdahl, et al. take this research one step further and analyze curricular efficiency and how the efficiency, or how streamlined a degree program is, affects on-time graduation rates [16]. The authors calculate curricular efficiency based on the minimum number of credit hours in the degree program, class prerequisite structure, course bottlenecks, the degree program's critical path, and the curricular rigidity, i.e., the total number of prerequisites and corequisites. However, much like the previously reviewed literature, the authors do not utilize AI tools, which can speed up this analysis process dramatically.

TABLE II. FACULTY COMMENTS ON SUGGESTED CURRICULAR CHANGES.

| Revision | Faculty Member One | Faculty Member Two |
|---|---|---|
| 1 | Suggested perhaps removing both prerequisites and only requiring that senior capstone be taken in a student's final year. Though they noted that the proposed action from the LLM was unclear | Noted that the SE 310 fail rate is relatively low and did not consider this to be an issue of importance. Also noted based on experience teaching senior capstone that the negative effect of missing SE 310 as a prerequisite on student performance in senior capstone would be too high |
| 2 | Agreed that CEC 470 may not be necessary and that important topics from that class are also covered in Real Time Systems (CEC 450) | Agreed that CEC 470 may not be necessary and that the latest version of ABET no longer requires it, but noted some concern that ABET could require it again in the future |
| 3 | They agreed but noted that the course number would have to change, which would affect the cross-listed graduate-level course | Agreed that this would be good, but not for the reason that Sonnet gave. Rather, students reach the quality assurance phase in senior capstone without having yet finished the Software Quality Assurance course, making that phase of their senior project difficult to complete |
| 4 | Expressed concern that the LLM was offering suggestions without considering the increase in credit load that these changes would cause | No comment. The faculty member thought this could be moved as needed to allow for necessary flexibility |
| 5 | Agreed, but noted that the department already did this | Agreed, but noted that the department already did this |

Finally, Heileman, et al. build on their own previous work to analyze degree program quality as a result of curricular complexity [2]. The authors use ANOVA statistical tests with the calculated complexities of randomly sampled top tier, mid tier, and bottom tier degree programs, according to the 2018 U.S. News & World Report, to determine the relationship. Their findings indicate that higher quality degree programs have lower curricular complexity.

### B. *Using LLMs to Modify Curricula*

Much of literature uses LLMs for generating syllabi or course learning objectives, but there is some existing work that utilizes LLMs for curricular analysis [17, 18]. Jayalath, et al. analyze the effectiveness of LLMs in comparison to previous ML-based methods for curriculum mapping, which analyzes whether curriculum meets desired outcomes [17]. Previously utilized ML-based methods include rule-based methods and Natural Language Processing (NLP) techniques including the use of BERT and other transformer models. The authors find that although LLMs offered stronger performance than other ML-based methods, they still struggled when skill outcomes were not explicitly defined in the provided documentation. These findings seem to align well with the findings in this paper, where inconsistencies or ambiguity within the provided documentation caused poorer performance. The work by Jayalath, et al., however, does not analyze curricular complexity which this work does.

Gacek and Adrian use LLMs to analyze curricular consistency, reduce topic redundancy, fill skill gaps, and fix sequencing errors [18]. The authors extract topics and skills from course syllabi using an LLM, construct a curriculum knowledge graph to structure the data for further logical inferencing, and use the knowledge graph to locate curricular inconsistencies. The paper therefore does not utilize the LLM for the curricular analysis which this work does.

Finally, Rutecka, et al. use LLMs to generate syllabi for undergraduate degree programs in Economics and Management [19]. The generated syllabi were compared to existing human-written syllabi for the courses and the authors found that while LLMs can likely aid humans in writing course syllabi, their performance indicates the need for heavy human involvement. Although the work does use LLMs for organizing and defining skill development within courses and curricula, it does not analyze curricular complexity which this work does.

## VI. Limitations and Future Work

Despite having the capabilities to maintain conversations in natural language, the majority of LLMs suffer from a great limitation, this being their inconsistent ability to process images correctly. When feeding the study plans to the LLMs, they were unable to preprocess the images correctly and understand the chains of courses that had to be taken sequentially. This caused the majority of the LLMs to fail to correctly identify the bottlenecks which could result in delayed graduation times.

Another problem that was encountered was the discrepancy between the online course catalog for the classes and the degree study plan flow charts when listing prerequisites for courses. Flow charts are only updated whenever significant changes are made to degree requirements such as adding new core classes, changing the credit hours certain classes are worth, or the number of credits required. Therefore, flowcharts do not contain the most up-to-date information. On the other hand, online course catalogs have the most correct listings of prerequisites for classes as they are updated as soon as changes happen.

One possible workaround would be to use Optical Character Recognition tools to convert the images to text before feeding those to the LLMs. This would allow LLMs to correctly understand the information the study plan graphs contain. This would potentially fix the issues understanding the study plan flow chart. Moreover, instead of using OCRs, Visual Language Models (VLMs) could be used to make the usage of the flow chart easier for the LLMs without the need for another tool that would increase the complexity of the process.

## VII. Conclusion

The purpose of this work was to identify whether LLMs could be used to analyze curricular complexity and if so, what the best LLM for this purpose is. This was achieved by comparing the analysis of five different LLMs, as well as one LLM using RAG on the curriculum for a B.S. in Software Engineering degree program. From this analysis, it is clear that Sonnet 4.6 without RAG is the best tool for this analysis as it was the only LLM to be able to correctly calculate the complexity factors. Upon calculating the complexity of the degree program, Sonnet 4.6 was also prompted for suggestions on how to reduce the complexity and improve on-time graduation rates. Based on feedback from Software Engineering professors, its suggestions were intriguing but could potentially cause some unintended consequences that would require further modeling and analysis. Therefore, Sonnet 4.6 could be a useful tool in calculating curricular complexity and brainstorming potential solutions for humans to then model and analyze further. Future work for this paper includes repeating the RAG experiment with a VLM and an OCR to improve the performance of the model in utilizing the degree program's flow chart.